\documentstyle[epsfig,aps,prl,bbm,twocolumn,amsmath,doublespace,12pt]{revtex}
\renewcommand{\theequation}{\arabic{equation}}
\begin{document}
%
%
%
%
\onecolumn
\title{The Majorana representation of spins and
the relation between  
$SU(\infty)$ and $SDiff(S^2)$}
\author{John Swain}
\address{Department of Physics, Northeastern University, Boston, MA 02115, USA\\
email: john.swain@cern.ch}
\date{April 29, 2004}
\maketitle

\begin{abstract}
\section*{\bf Abstract}
The Majorana representation of spin-$\frac{n}{2}$ quantum states by sets of 
points on a sphere allows a realization of $SU(n)$ acting on 
such states, and thus a natural action on the two-dimensional sphere $S^2$.
This action is discussed in the context of 
the proposed connection between
$SU(\infty)$ and the group $SDiff(S^2)$ of area-preserving diffeomorphisms
of the sphere. There is no need to work with a special basis of the 
Lie algebra of $SU(n)$, and there is a clear geometrical interpretation of
the connection between the two groups. It is argued that 
they are {\it not} isomorphic, and comments are made concerning the 
validity of approximating groups of area-preserving diffeomorphisms by 
$SU(n)$. 

\end{abstract}

\section{Introduction}

Groups of area-preserving diffeomorphisms and their Lie algebras
have recently been the focus of much attention in the physics 
literature.    
Hoppe \cite{Hoppe1} has shown that in a suitable basis, the Lie algebra of 
the group $SDiff(S^2)$ of area-preserving diffeomorphisms of a
sphere tends to that
of $SU(N)$ as $N\rightarrow\infty$.
This has obvious interest in 
connection with gauge theories of $SU(N)$ for large N. 
The use of $SU(N)$ for finite $N$ as
an approximation to the group of area-preserving diffeomorphisms has also been
used in studies of supermembranes \cite{supermem1,supermem2,supermem3} and 
in particular has been used to argue for their instability.

The limiting procedure as $N\rightarrow\infty$ is delicate, and in particular, 
the need to take the limit in a particular basis makes one immediately 
wary as to how this result should be interpreted. In fact Hoppe and
Schaller \cite{Hoppe2}
have shown that there are infinitely many pairwise non-isomorphic Lie algebras,
each of which tends to $su(\infty)$, the Lie algebra of $SU(\infty)$,
as $N\rightarrow\infty$.
The authors of references \cite{supermem2}, \cite{supermem3}
and \cite{mylimits,topology}
have especially
emphasized the difficulties in relating such infinite limits with Lie algebras
of area-preserving diffeomorphisms.
Various authors
have considered special limits and/or large-N limits of
other classical Lie algebras \cite{Pope,Pope2,Wolski,deWit,Vassilevich}
as relevant for 2-manifolds other than spheres. 
The purpose of this Letter is to clarify the geometrical nature of
the relationship between $SU(N)$ and $SDiff(S^2)$.

In this Letter we will consider the group $SU(N)$ for $N\rightarrow\infty$
without the use of a specific basis for its Lie algebra, and in fact without
consideration of its Lie algebra at all! The argument requires some
familiarity with the Majorana representation of a spin-$j$ system by a 
set of $2j$ points, not necessarily all distinct,
on the surface of a 2-dimensional sphere, $S^2$. 

\section{The Majorana Representation of Spin}

Note that a system with classical angular momentum $\vec{J}$ can be described
by a single point on $S^2$ corresponding to the direction in which
$\vec{J}$ points.
The case in quantum mechanics is more subtle. A general state of
spin $j$ (in units
of $\hbar$) must be represented by a collection of $2j$ points on the surface
of a sphere, as first shown by Majorana \cite{Majorana} and later by
Bacry \cite{Bacry}
(for a summary of both papers, see reference \cite{Biedenharn}).

We reproduce the argument here for completeness, as it provides a
connection between the action
of $SU(2j)$ on the complex projective Hilbert space
${\mathbbm{C}}^{2j}$ representing
states of spin $j$ and diffeomorphisms of $S^2$. 

Let ${\mathbbm{CP}}^{2j}$ denote the projective space associated with 
${\mathbbm{C}}^{2j+1}$. This is the space of $2j+1$-tuples
in ${\mathbbm{C}}^{2j+1}$ considered 
equivalent if they differ by a complex scale factor $\lambda$. That is,
two points $(a_1,a_2,...,a_{2j+1})$, and
$(\lambda a_1,\lambda a_2,...,\lambda a_{2j+1})$
of ${\mathbbm{C}}^{2j+1}$
are considered the same point of ${\mathbbm{CP}}^{2j}$. 

Now consider the set $P_{2j}$ of nonzero homogeneous polynomials of 
degree $2j$ in two complex variables, $x$, and $y$, which we associate
to $(2j+1)$-tuples as follows:

\begin{equation}
(a_1,a_2,...,a_{2j+1}) 
\rightarrow a_1 x^{2j} + a_2 x^{2j-1} y + \cdots + a_{2j+1}y^{2j}
\end{equation}

Now (by the fundamental theorem of algebra) the polynomials can be factored 
into a product of $2j$ (not necessarily different) homogeneous terms
linear in $x$ and
$y$. For each 
polynomial in $P_{2j}$ we can associate an element of ${\mathbbm{C}}^{2j+1}$
by writing it as a product of 2$j$ factors:

\begin{equation}\label{eq:factors}
a_1 x^{2j} + a_2 x^{2j-1} y + \cdots + a_{2j+1}y^{2j} = (\alpha_1 x_1 -\beta_1
y_1)(\alpha_2 x_2 - \beta_2 y_2) \cdots (\alpha_n x_n - \beta_{2j+1} y_{2j+1})  
\end{equation}

\noindent for some complex $\alpha_i,\beta_i$.

The coefficients of $x$ and $y$ in each term represent
a point in ${\mathbbm{C}}^{2j+1}$.
This factorization
is not unique, in that the terms can be permuted, and one can multiply
any two factors by $\alpha$ and $\alpha^{-1}$ respectively. Thus we see
that the whole space ${\mathbbm{CP}}^{2j}$ in one-to-one correspondence with
unordered sets of $2j$ points in ${\mathbbm{CP}}^1$ corresponding to the
$2j$ terms in equation \eqref{eq:factors} where we identify
$(x_i,y_i) = \lambda (x_i,y_i)$ for any $\lambda$.
But ${\mathbbm{CP}}^1$ is $S^2$
and so we have the required result : {\it States of angular momentum $2j$
can be represented by single points in ${\mathbbm{CP}}^{2j}$ or unordered
sets of $2j$ 
(not necessarily distinct) points on $S^2$}.

Another, though less explicit, way to understand the Majorana 
representation is to note that a state of spin-$j$ can be
written as a totally symmetric product of $2j$ spin-1/2
wavefunctions.

\section{$SU(N)$ and its action on spin states}

Recall that ${\mathbbm{CP}}^n$ is the set of lines
in ${\mathbbm{C}}^{n+1}$ and
can be written as the coset space

\begin{equation}
{\mathbbm{CP}}^n = SU(n+1)/U(n)
\end{equation}
                   
Now for any coset space $S = G/H$, G acts transitively
on $S$ (that is, for any two points $p$ and $q$ of
$S$, there is an element of $G$ which takes $p$ to 
$q$, so that $Gp=q$). Thus $SU(2n+1)$ has a natural 
action on ${\mathbbm{CP}}^{2n}$, and thus on sets of points of
$S^2$ such that any unordered set of points is carried
to any other unordered set of points by a suitable 
transformation from $SU(2n+1)$.

So far, what has been presented is valid for any finite $n$.
For each finite $n$ then, we have a realization of the
action of $SU(n)$ on $S^2$.

Now with this action on $S^2$,
$SU(n)$ will {\it always} (even for finite $n$) 
contain transformations which carry two distinct points
into the same point on
$S^2$, and thus which will not correspond to 
diffeomorphisms of $S^2$. (We note in passing that $SDiff(S^2)$
is k-fold transitive for every positive integer k \cite{Boothby}
where we recall that the action of a group $G$ on a manifold $M$
is said to be k-fold transitive if for any two arbitrary sets
of k {\it distinct} points $(p_1,p_2,p_3,\ldots,p_k)$ and
$(q_1,q_2,q_3,\ldots,q_k)$ of $M$ there is an element of $G$ which takes
$p_i$ to $q_i$ for all $i=1,\ldots,k$).

Thus in the limit of very large $n$, the permutations of
sets of points on the sphere become much larger than
the set of all diffeomorphisms of the sphere.
In particular, any
finite $N$ approximation, $SU(N)$ of $SDiff(S^2)$
will contain mappings which do
not correspond to elements of $SDiff(S^2)$.
Thus we see that

\begin{equation}
\lim_{N\rightarrow\infty} SU(N) \not\simeq SDiff(S^2)
\end{equation}

This can also be seen from another geometric
view, where the Lie algebra of $SDiff(S^2)$ is that of divergenceless
vector fields on $S^2$. Clearly, points pushed along these integral
curves of these vector
fields must not meet, and yet we see that there exist elements of
$SU(N)$ acting on $S^2$ which will not satisfy this requirement.
Of course one can argue that in some sense ``most'' of the transformations
will not
carry two distinct points into one, and in fact this argument is 
implicit in the assumptions about the limiting procedures which
associate the Lie algebras of $SDiff(M)$ and of $SU(\infty)$ for
various choices of 2-manifold $M$ \cite{supermem3}, but this does
not evade the fact that $SU(\infty)$ contains transformations
which are clearly not in $SDiff(M)$.

It is interesting to note that $SU(\infty)$ as realized here, is, in fact, 
so large, and capable of such dramatic topology-changing distortions of
$S^2$ that it may in fact be a useful description not of 
area-preserving diffeomorphisms of $S^2$, but rather of 
a wider class of deformations of $S^2$ including those which 
result in punctured spheres, 2-manifolds of different
topologies, and 2-manifolds which have degenerated into
1-manifolds, or even a single point. Similar ideas have been
put forth in \cite{supermem3}.

\section{Acknowledgement}

The author would like to
my colleagues at Northeastern University and the National Science
Foundation for their support.

\end{document}